\documentclass[sigconf, screen]{acmart}
\usepackage[utf8]{inputenc}
\usepackage{amsmath}
\usepackage{algorithmic}
\usepackage{graphicx}
\usepackage{textcomp}
\usepackage{color}
\usepackage{float}
\usepackage{fontenc}
\usepackage[toc,page]{appendix}

\setcopyright{acmcopyright}
\copyrightyear{2021}
\acmYear{2021}
\acmDOI{10.1145/1122445.1122456}

\acmConference[CARRV 2021]{CARRV 2021: Fifth Workshop on Computer Architecture Research with RISC-V}{June 17, 2021}{Virtual}
\acmBooktitle{CARRV 2021: Fifth Workshop on Computer Architecture Research with RISC-V,
  June 17, 2021, Virtual}
\acmPrice{15.00}
\acmISBN{978-1-4503-XXXX-X/18/06}

\begin{document}

\title{Arrow: A RISC-V Vector Accelerator for Machine Learning Inference}
\author{Imad Al Assir, Mohamad El Iskandarani, Hadi Rayan Al Sandid, and Mazen A. R. Saghir}
\email{{iha14,mbe22,hxa04}@mail.aub.edu, mazen@aub.edu.lb}
\affiliation{%
  \institution{~\\Embedded and Reconfigurable Computing Lab\\Department of Electrical and Computer Engineering\\American University of Beirut}
  \city{Beirut}
  \country{Lebanon}
  }

\renewcommand{\shortauthors}{Al Assir, et al.}

\begin{abstract}

In this paper we present Arrow, a configurable hardware accelerator architecture that implements a subset of the RISC-V v0.9 vector ISA extension aimed at edge machine learning inference. Our experimental results show that an Arrow co-processor can execute a suite of vector and matrix benchmarks fundamental to machine learning inference 2 -- 78$\times$ faster than a scalar RISC processor while consuming 20\% -- 99\% less energy when implemented in a~Xilinx XC7A200T-1SBG484C FPGA.
\end{abstract}

\keywords{Computer architecture, vector accelerator, RISC-V, FPGA, machine learning inference}

\maketitle

\section{Introduction}

The proliferation of computationally demanding machine learning applications comes at a time when traditional approaches to processor performance are providing diminishing returns. This has renewed interest in domain-specific architectures and hardware application accelerators as a means for achieving the required levels of performance~\cite{goldenage}. The RISC-V instruction-set architecture (ISA), with its rich set of standard extensions and ease of customization, has become the ISA of choice for innovations in domain-specific processor design. In particular, the RISC-V vector extensions are a~good match for workloads such as machine learning inference that exhibit high levels of data parallelism. Vector architectures also provide a low-energy processing alternative for energy-limited applications such as edge computing. 

In this paper we present our Arrow accelerator architecture and report on its performance and energy consumption on a suite of vector and matrix benchmarks that are fundamental to machine learning inference. In Section~\ref{sec:rel_work} we describe related work and show how the use of vector architectures has evolved over time. In Section~\ref{sec:arch} we describe the Arrow microarchitecture and the various design decisions that we made. In Section~\ref{sec:method} we describe our experimental methodology and the tools we used to implement our design and evaluate its performance. In Section~\ref{sec:results} we present our experimental results, and in Section~\ref{sec:conc} we present our conclusions and describe future work.

    \begin{figure*}[ht!]
        \centering
        \includegraphics[width=\textwidth]{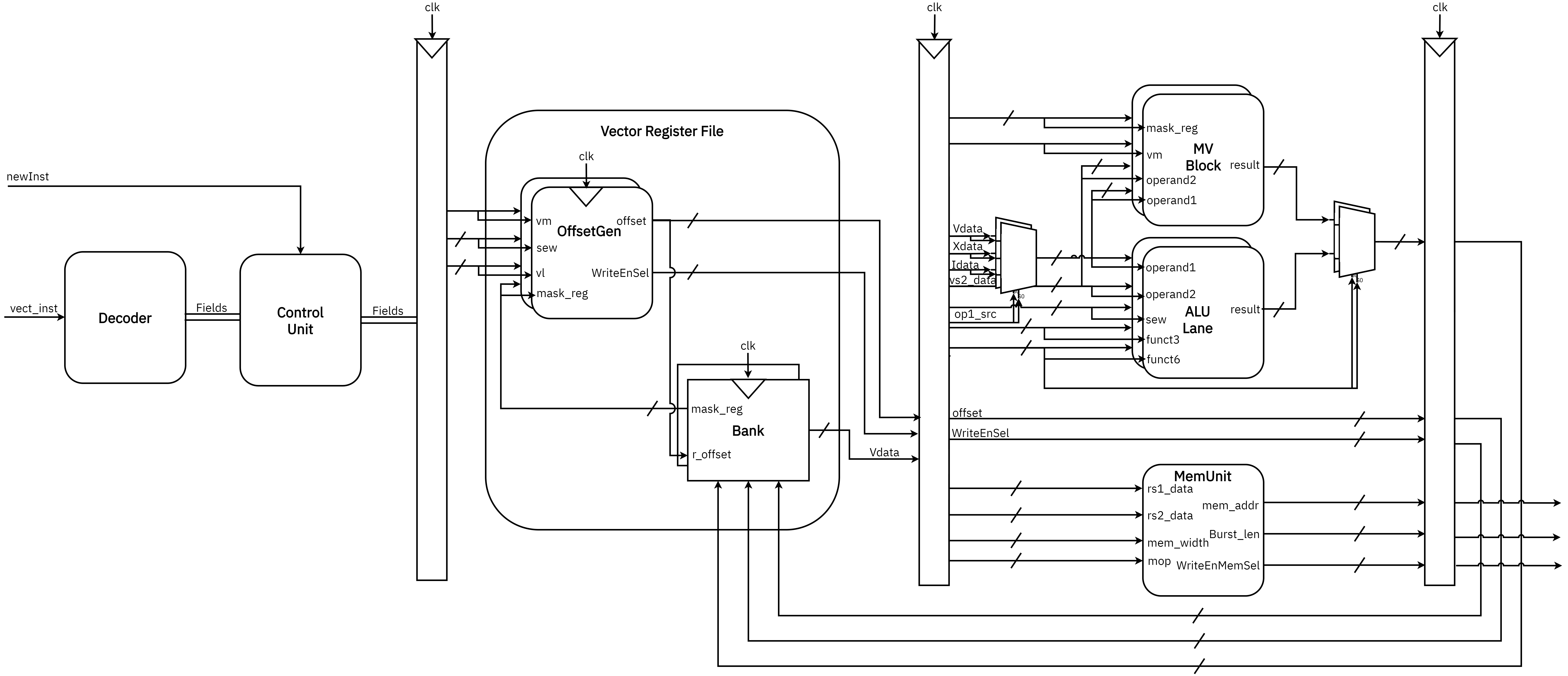}
        \caption{Arrow Datapath}
        \label{fig:datapath}
    \end{figure*}

\section{Related Work} \label{sec:rel_work}

The earliest vector processors were developed in the mid 1970's. They were used in supercomputers like the Cray-1~\cite{cray1} to run scientific applications with high levels of data parallelism. These processors held a performance advantage over microprocessors of the time due to their faster clock rates and their architectures, which achieved higher computational and memory throughput~\cite{valero98}. 

By the early 1990's, fast and inexpensive ``killer micros" like the DEC Alpha 21164~\cite{alpha} became an attractive alternative for building scalable parallel computer systems. These ultimately displaced vector machines in high-performance computing systems~\cite{valero98}.

The early 1990's also saw increased demand for efficient architectures aimed at data-intensive digital signal processing, graphics, and multimedia applications. Microprocessor manufacturers responded by extending their ISAs to support short-vector and SIMD instructions. ISA extensions such as Intel's MMX/SSE and ARM's Neon are still in use, but some researchers argue that they lead to instruction-set bloat and increase processor complexity~\cite{patterson17}.

From the mid 1990's to the early 2000's, academic interest in single-chip vector processors led to the design of the Berkeley T0  and VIRAM processors~\cite{asanovic98,viram03}. The T0 processor demonstrated the viability of integrating a scalar MIPS processor with a~heterogeneous, chained, multi-lane vector co-processor. The VIRAM processor demonstrated the superiority of vector architectures over superscalar and VLIW architectures for data-intensive embedded applications. Its vector co-processor was designed around homogeneous, non-chained, vector lanes and distributed memory banks.

The late 2000's also saw academic interest in using FPGAs to implement scalable soft vector processors (SVPs) inspired by the VIRAM architecture.  VESPA~\cite{vespa08} was a MIPS-based SVP with up to 16 homogeneous vector lanes. VESPA was later enhanced to support heterogeneous vector lanes, chaining, and a banked vector register file~\cite{vespa2}. VIPERS~\cite{vipers} was another VIRAM-inspired SVP that was designed as a general-purpose accelerator. It included a configurable multi-lane datapath and memory interface. Later SVPs introduced  new features like using a scratchpad memory as a vector register file~\cite{vegas}, and implementing 2D and 3D vector instructions~\cite{venice}. Most of these features are incorporated in at least one commercial SVP~\cite{vectorblox}.

The emergence of the RISC-V ISA in 2010 coincided with the proliferation of deep machine learning algorithms and renewed interest in accelerating machine learning training and inference using domain-specific architectures. Berkeley's Hwacha vector architecture~\cite{lee16} inspired the development of a standard RISC-V vector extension (RVV)~\cite{rvv09}. Today, a number of commercial products and academic projects are based on implementations of the RVV ISA extension. 

The Andes NX27V and SiFive VIS7~\cite{andes, sifive} are commercial processors based on the RVV v1.0 extension. They can operate on 512-bit vectors and are aimed at data-intensive applications in telecommunications, image/video processing, and machine learning. ETH Zurich's Ara is a vector co-processor for the 64-bit Ariane RISC-V processor that is based on the RVV v0.5 ISA extension~\cite{cavalcante2019ara}. It is a~scalable architecture that was implemented in 22 nm FD-SOI technology and could operate at 1 GHz+. Four instances of the co-processor were implemented with 2, 4, 8, and 16 lanes, respectively. Performance results demonstrated the efficiency of the architecture on data-parallel workloads. Ara's support for double-precision floating-point operations makes it well suited for HPC-class data-parallel applications. The University of Southampton's AVA is a RVV co-processor for the OpenHW Group CV32E40P that is optimized for machine learning inference~\cite{Southampton_AVA}. It implements a subset of the RVV v0.8 ISA extension, uses 32, 32-bit vector registers, and can operate on 8-, 16-, or 32-bit vector elements in SIMD fashion. The AVA co-processor is implemented in Verilog and was validated through RTL and instruction set simulation. An earlier version of the co-processor targeting embedded applications was implemented in an Intel Cyclone V FPGA and could operate at 50~MHz~\cite{USouthamptonRVVEmbedded2020}. Finally, Sapienza University of Rome's Klessydra-T13~\cite{klessydra} is a RISC-V multi-threaded processor with a~configurable vector unit. It uses custom vector instructions that replicate some of the basic operations of the standard RVV extension and implement specialized functions for machine learning workloads.

\section{Arrow Architecture} \label{sec:arch}

Arrow is a configurable vector co-processor that implements a~subset of the RVV v0.9 ISA extension. Some of its architectural parameters can be configured at design time including the number of lanes, maximum vector length ($VLEN$), and maximum vector element width ($ELEN$). In this paper we report on a dual-lane Arrow architecture with $VLEN$ = 256 bits and $ELEN$ = 64 bits. The current implementation does not support chaining.

\subsection{Arrow RISC-V Vector ISA Subset}

The RISC-V vector extension provides a rich set of vector instructions. 
Arrow implements a subset of these instructions that we believe can be useful in edge machine learning inference applications. 
These include unit-stride and strided memory access; single-width integer addition, subtraction, multiplication, and division; bitwise logic and shift; and integer compare, min/max, merge, and move operations.

\subsection{Pipeline Organization}

Figure \ref{fig:datapath} shows Arrow's datapath.  
It is a single-issue, dual-lane vector accelerator that can execute two independent vector instructions in parallel. It is pipelined to increase execution throughput, and its pipeline stages include decode, operand fetch, execute or memory access, and write-back. 

Instructions are dispatched from a scalar host processor and are first decoded. Different instruction fields are extracted and fed into the control unit, which generates the appropriate control signals. An offset generator calculates the vector register offsets needed to access the corresponding vector elements. These offsets are sent to the vector register file banks to read the required data. Operands are selected based on the control signals, and depending on the instruction type, the required fields are fed to the ALU, move block, or memory unit. For vector load instructions, the appropriate vector elements are passed to the memory interface, and for arithmetic or logical instructions, results are written back to the register file. The move block executes regular and masked merge and move instructions. 

\subsection{Decoder and Controller}

The decoder is a combinational circuit that decomposes the 32-bit instruction it receives from the scalar processor into the appropriate fields needed by other datapath components. The  controller uses these fields to generate the appropriate control signals for each vector lane. It also passes control signals and dispatches instructions for execution in the appropriate lane based on the destination vector register. For our dual-lane architecture, if the destination register corresponds to registers 0 to 15 the vector instruction is executed in the first lane. Similarly, if the destination register corresponds to registers 16 to 31 the instruction is executed in the second lane. This lane-dispatching scheme simplifies our design and eliminates the need for complex arbitration hardware. It also supports wider lane implementations, but requires the compiler or assembly programmer to expose vector instruction parallelism through register allocation in a manner similar to statically scheduled superscalar processors. Once generated, control signals are held constant for the duration of a vector instruction execution.

\subsection{Vector Register File}

Arrow's register file is based on a banked architecture where each bank is associated with a different lane. In the current implementation, the first bank contains registers 0 to 15 while the second bank contains registers 16 to 31. Each bank has two read ports and one write port. This enables both banks to support the two lanes simultaneously.

\begin{figure}[ht]
\centering
\includegraphics[width=\columnwidth]{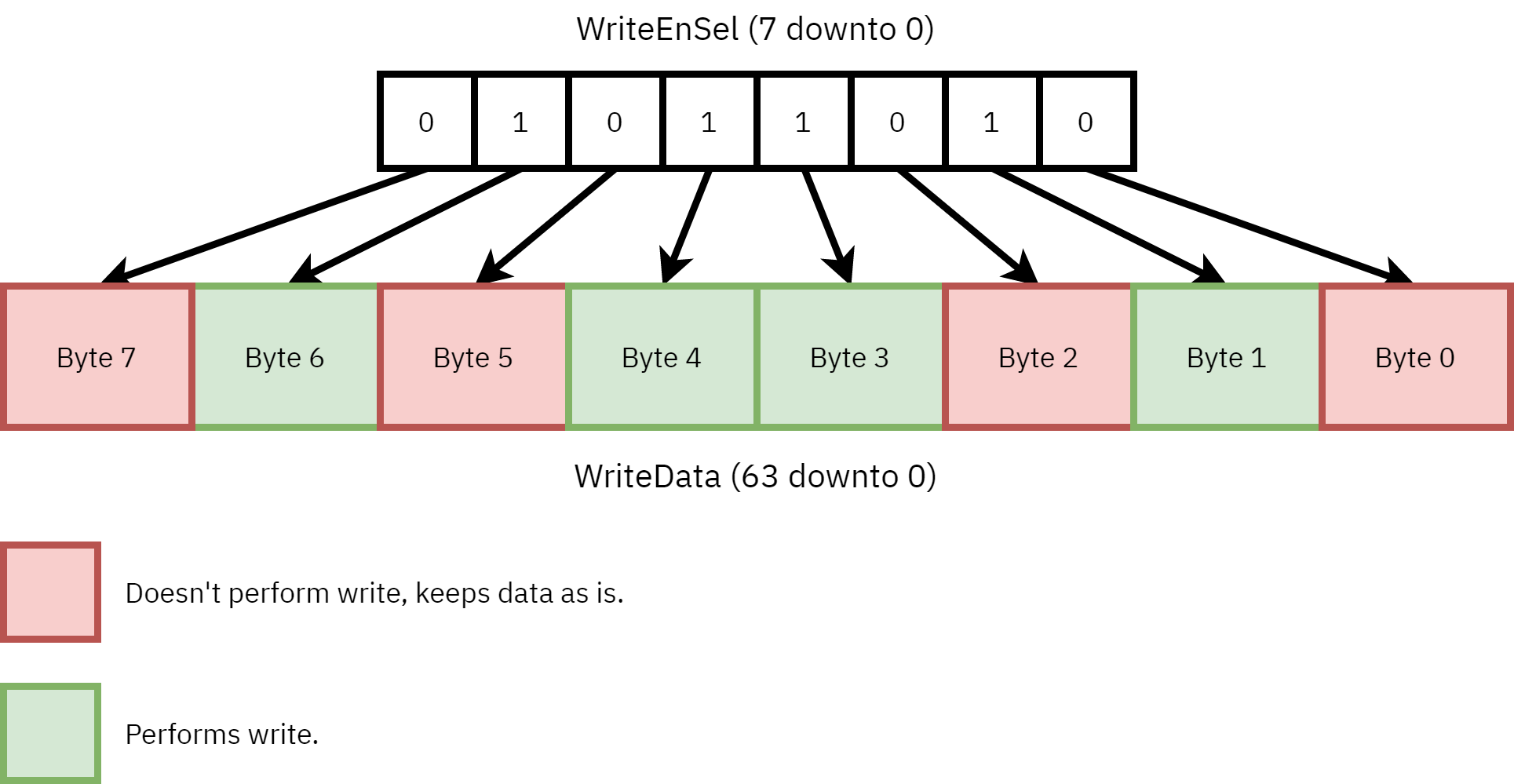}
\caption{The WriteEnable bits mapping }
\label{fig:WriteEn}
\end{figure}

Within Arrow, vector data is processed in $ELEN$-bit words. An offset generator associated with each register bank is used to generate the byte offsets for accessing the corresponding $ELEN$-bit words from a vector register. The offset generator also generates a write-enable selector mask that specifies which bytes within an $ELEN$-bit word should be updated when a result is written back to a vector register. The number of bytes associated with a vector element is determined by the $SEW$ parameter, and $\lceil\text{$VLEN$/$ELEN$}\rceil$ offsets are generated for each vector register. Figure \ref{fig:WriteEn} shows how the write-enable selector bits can be used to mask the write back to arbitrary bytes in an $ELEN$-bit word.

        
\subsection{SIMD ALU}

\begin{figure}[h]
\centering
\includegraphics[width=\columnwidth]{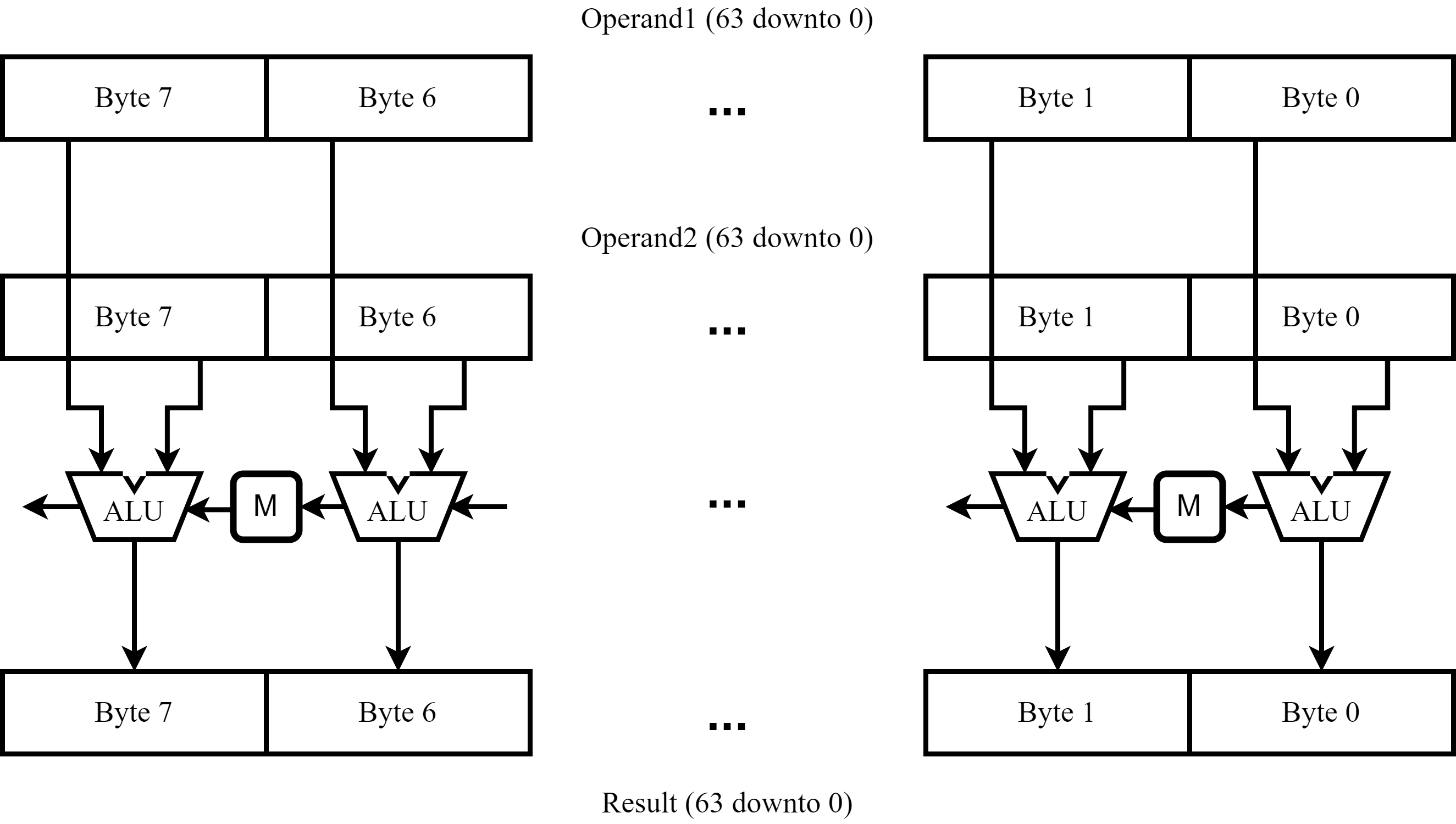}
\caption{SIMD ALU Design}
\label{fig:SIMD}
\end{figure}

The Arrow ALU operates on $ELEN$-bit words regardless of the actual standard element width ($SEW$) being used. When $SEW$ is less than $ELEN$, the ALU processes multiple vector elements simultaneously in SIMD fashion. Figure \ref{fig:SIMD} shows the Arrow SIMD ALU, which performs basic arithmetic and logic operations. Depending on the actual size of a vector element ($SEW$), multiplexers (denoted by M in the Figure) ensure proper carry chain propagation.

\subsection{Memory Unit} 

The RVV ISA extension supports different vector memory addressing modes including unit stride, strided, and indexed access. The Arrow memory unit is designed to support these addressing modes by generating corresponding memory addresses. When executing a vector memory instruction, the base address is received from the scalar processor through the \texttt{rs1\_data} port.
The memory unit then generates the corresponding effective memory addresses and burst length. The latter is used to determine the number of $ELEN$-bit words that need to be transferred from memory, thus allowing efficient multi-beat bursts. For vector load instructions, the memory unit also generates the \texttt{WriteEnMemSel} bit mask to indicate which vector element bytes should be written in the vector register file. Currently, both unit-stride and strided accesses are fully supported, but vector indexed/gather-scatter access is still in development.

\begin{figure*}[ht!]
\centering
\includegraphics[width=\textwidth]{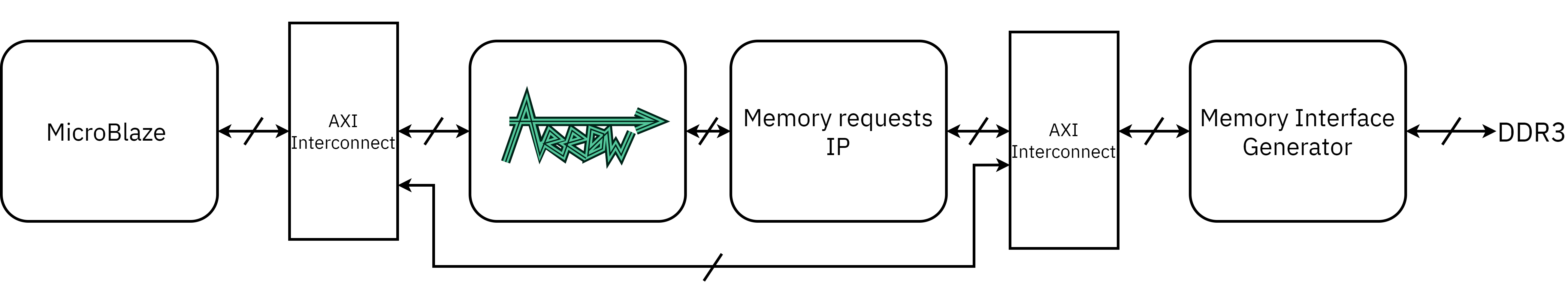}
\caption{FPGA system block diagram. System bus is $ELEN$=64-bits wide.}
\label{fig:FPGA_design}
\end{figure*} 

\subsection{Memory Bus Interface}

Because we are using Xilinx design tools, our Arrow system architecture is based on the
ARM AMBA AXI bus interface and AXI IPs~\cite{armAMBAAXIACE}. Figure~\ref{fig:FPGA_design} shows a high-level view of the memory interface, which includes the Arrow memory unit, Arrow AXI master interface, Xilinx Memory Interface Generator (MIG) IP block, and DDR3 memory. The DDR3 memory provides a shared address space for both the MicroBlaze and Arrow processors. Our system does not currently use any cache or scratchpad memories.

Within Arrow, the memory unit generates the addresses of required $ELEN$-bit memory words. It also determines the number of memory words that need to be transferred -- we refer to these as the number of beats or the burst length. In our current implementation, all memory accesses are $ELEN$=64 bits wide regardless of whether the entire data are needed or not. This simplifies the memory interface and avoids narrow transactions that are smaller than the AXI bus width. For strided vector load instructions, the memory unit also generates the masks that enable the writing of specific bytes to the vector register file. Arrow's AXI master interface is shown as the Memory Requests IP block in Figure~\ref{fig:FPGA_design}. It translates an Arrow memory request to an AXI bus transaction. The MIG IP block is external to the Arrow architecture. It serves as a DDR3 memory controller that converts Arrow memory transactions over the AXI bus to DDR3 memory transactions. It also provides the MicroBlaze processor with access to DDR3 memory.

The Arrow memory interface is designed to be system-bus independent. However, in the current implementation, some limitations of the MIG interface can impact performance. For example, the MIG interface is limited to 64-bit transfers. Although this matches Arrow's internal $ELEN$-bit word length, the MIG interface does not support concurrent or interleaved AXI memory transfers. This limits vector memory transfers on the Arrow side to a single lane, which can increase latency and reduce throughput. On the other hand, because the 16-bit MIG/DDR3 interface can operate at 400~MHz, roughly four times the speed of the Arrow core, we can read or write an $ELEN$-bit word every AXI bus cycle.

\section{Experimental Methodology} \label{sec:method}

In this section we describe our experimental methodology, which includes our FPGA prototype, the code development tools we used, and the fundamental machine learning vector and matrix benchmarks we used to evaluate performance.

\subsection{FPGA Prototype}
\label{sec:FPGA_proto}
After developing the Arrow microarchitecture in VHDL, we synthesized and implemented our design in the Xilinx XC7A200T-1SBG484C FPGA used in the Nexys Video board \cite{nexysvideo}. We also used the Xilinx Vivado 2019.1 Design Suite for synthesis and implementation. 

Figure \ref{fig:FPGA_design} shows our FPGA system implementation. The Arrow datapath is packaged as an AXI IP to enable communication with the MicroBlaze processor and memory over the AXI bus. We chose the Xilinx MicroBlaze v11.0 \cite{MicroBlaze} as our scalar processor because of its compatibility with the Vivado 2019.1 design suite and AXI bus interface. This enabled us to run C test code and send Arrow instructions through the AXI bus. We also validated Arrow's internal signals using the Integrated Logic Analyzer (ILA) IP block, which is available through the Vivado Design Suite. We also developed a standalone IP block that translates Arrow memory requests to AXI bus transactions to ensure the Arrow IP block remained as generic and bus-independent as possible. The DDR3 memory and MIG interface are shared by both the MicroBlaze and Arrow processors.

\subsection{Code Development Tools}

To help us cross-compile benchmark programs written in C/C++, we used the RISC-V LLVM/Clang toolchain from the European Processor Initiative (EPI) project~\cite{EPI-llvm}, which supported the RVV v0.9 ISA extension specification. Although LLVM/Clang supports the use of RVV instrinsics, we used assembly code inlining to have greater control over our code. We then packaged the inlined code in functions that we used in our benchmarks.

To validate the functionality of our benchmarks, we used the open-source Spike RISC-V ISA simulator~\cite{RiscvRiscvisasim2021}. In addition to supporting the base RISC-V instruction set, Spike supports most of the standard ISA extensions including the "V" vector extension. However, because Spike is not cycle-accurate~\cite{spike}, we developed our own cycle count models to evaluate and compare the execution performance of both the scalar and vector benchmarks. To validate our cycle count model, we compared its cycle counts for the scalar benchmark to those produced by Spike and found them to be within~7\%.

\subsection{Benchmarks}

To evaluate our design, we used a suite of vector and matrix benchmarks from the University of Southampton~\cite{AVA_benchmarks}. We chose these benchmarks because they are widely used in machine learning inference algorithms, and because they exhibit high levels of data parallelism. The benchmarks include vector  addition,  multiplication, max reduction, dot product, and rectified linear unit (ReLu). They also include matrix multiplication, addition, max pool, and 2D convolution. 

The original benchmark suite was developed for the RVV v0.8 ISA extension specification. To support the RVV v0.9 specification, we made minor adjustments to its inlined assembly code functions. We then compiled the applications using LLVM/Clang and simulated them on Spike for functional validation. Finally, we compared the scalar and vector execution times by applying our cycle count models. We also calculated the energy used by each benchmark by multiplying the power consumed, which we obtained from the FPGA synthesis reports, by the execution time. We calculated a~benchmark's execution time by multiplying its cycle count by the Arrow clock cycle time.  To study the impact of data size on performance and energy consumption, we also ran our benchmarks with small, medium and large data sizes. Table~\ref{tab:presets} describes our different data size profiles.

\begin{table}
    \centering
    \small
    \begin{tabular}{|l|c|c|c|}
    \hline
    & Small Data & Medium Data & Large Data \\
    & Profile & Profile & Profile \\ \hline
     Vector Length & 64 elements & 512 elements & 4096 elements\\ \hline
     Matrix Size & $64\times64$ & $512\times512$ & $4096\times4096$ \\  \hline
     2D Convolution & & & \\
     Data Size & $1024\times1024$ & $1024\times1024$ & $1024\times1024$\\
     Kernel Size & $3\times3$ & $4\times4$ & $5\times5$\\
     Channels & 1 & 1 & 1 \\
     Batch Size & 3 & 4 & 5\\ \hline
    \end{tabular}
    \caption{Benchmark Parameters}
    \label{tab:presets}
\end{table}

\section{Results} \label{sec:results}

In this section, we report on the results of implementing the Arrow system in a FPGA device. We also report on the performance and energy of the Arrow accelerator on vector and matrix benchmarks.

\subsection{FPGA Resource Utilization}
To evaluate our design, we synthesized and implemented two systems: one with a single MicroBlaze processor and another with both MicroBlaze and Arrow processors. Both systems ran at 100~MHz. Table~\ref{tab:mb_res} summarizes our post-implementation results, which demonstrate the minor increase in LUT and FF utilization, and power consumption, due to the Arrow accelerator. And although we operated Arrow at 100~MHz, its maximum clock speed is 112~MHz.

\begin{table*}[ht]
    \centering
    \begin{tabular}{||c||c||c||c||c||}
    \hline \hline
         & \multicolumn{3}{c||}{Utilization} & Power\\
        \cline{2-4} 
         & LUT & FF & BRAM &  (W)\\
         \hline
         MicroBlaze & 2241/133800 (1.7\%) & 1495/267600 & 32/365 & 0.270 \\ \hline
         MicroBlaze+Arrow & 2715/133800 (2.0\%) & 2268/267600 & 32/365 & 0.297 \\
         \hline\hline
    \end{tabular}
    \caption{FPGA Implementation Results}
    \label{tab:mb_res}
\end{table*}

\subsection{Benchmark Performance}

\begin{table*}[ht!]
 \centering

\scalebox{0.8}{
\begin{tabular}{@{\extracolsep{2pt}}||l||c|c|c||c|c|c||c|c|c||}
\hline\hline
 {}  & \multicolumn{3}{c||}{Small Data Profile}  & \multicolumn{3}{c||}{Medium Data Profile} & \multicolumn{3}{c||}{Large Data Profile} \\
 \cline{2-4} 
 \cline{5-7} 
 \cline{8-10} 
 Operation  & Scalar (cycles) & Vector (cycles) & Speedup & Scalar (cycles) & Vector (cycles) & Speedup & Scalar (cycles) & Vector (cycles) & Speedup \\ 
\hline
Vector Addition    & $3.4\times{}10^{3}$ & $5.0\times{}10^{1}$  & $69.6\times$ & $2.7\times{}10^{4}$  & $3.5\times{}10^{2}$  &  $77.3\times$ & $2.2\times{}10^{5}$   & $2.8\times{}10^{3}$  &  $78.4\times$  \\ \hline
Vector Multiplication    & $3.5\times{}10^{3}$  & $5.0\times{}10^{1}$ & $69.5\times$ & $2.8\times{}10^{4}$ & $3.6\times{}10^{2}$ &  $77.3\times$  & $2.2\times{}10^{5}$ & $2.8\times{}10^{3}$ &  $78.3\times$  \\ \hline

Vector Dot Product    & $1.6\times{}10^{3}$  & $6.2\times{}10^{1}$ & $25.2\times$ & $1.2\times{}10^{4}$ & $3.8\times{}10^{2}$ & $32.1\times$  & $9.8\times{}10^{4}$ & $3.0\times{}10^{3}$ & $33.2\times$  \\ \hline

Vector Max Reduction & $1.4\times{}10^{3}$  & $4.2\times{}10^{1}$ & $32.6\times$ & $1.1\times{}10^{4}$ & $2.2\times{}10^{2}$ & $48.1\times$ & $8.6\times{}10^{4}$ & $1.7\times{}10^{3}$ & $51.2\times$    \\ \hline

Vector ReLu & $1.4\times{}10^{3}$  & $4.2\times{}10^{1}$ & $34.0\times$ & $1.1\times{}10^{4}$ & $2.9\times{}10^{2}$ & $38.4\times$ & $9.0\times{}10^{4}$ & $2.3\times{}10^{3}$ & $39.0\times$    \\ \hline

Matrix Addition  & $2.2\times{}10^{4}$  & $5.1\times{}10^{3}$ & $43.8\times$ & $1.4\times{}10^{7}$ & $2.0\times{}10^{5}$ & $71.6\times$ & $9.1\times{}10^{8}$ & $1.2\times{}10^{7}$ & $77.6\times$   \\ \hline

Matrix Multiplication  & $1.2\times{}10^{7}$  & $5.1\times{}10^{5}$ & $24.1\times$ & $6.1\times{}10^{9}$ & $1.2\times{}10^{8}$ & $50.4\times$ & $3.1\times{}10^{12}$ & $5.3\times{}10^{10}$ & $58.6\times$    \\ \hline

 Matrix Max Pool & $3.7\times{}10^{5}$  & $7.0\times{}10^{4}$ & $5.4\times$ & $2.4\times{}10^{7}$ & $4.4\times{}10^{6}$ & $5.4\times$ & $1.5\times{}10^{9}$ & $2.8\times{}10^{8}$ & $5.4\times$  \\ \hline
 
 2D Convolution  & $1.4\times{}10^{9}$  & $7.3\times{}10^{8}$ & $1.9\times$ & $1.9\times{}10^{9}$ & $1.2\times{}10^{9}$ & $1.6\times$  & $2.4\times{}10^{9}$ & $1.8\times{}10^{9}$ & $1.4\times$    \\ 
\hline\hline
\end{tabular}}
\caption{Cycle-Count Performance Analysis}
\label{tab:Cycle_count}
\end{table*}

Table~\ref{tab:Cycle_count} shows the cycle counts for the scalar and vectorized benchmarks for different data size profiles. The Table also shows the speedup factors of vectorized benchmarks relative to scalar benchmarks. Our results show that the vector benchmarks run \mbox{25 -- 78$\times$} faster; the matrix benchmarks run \mbox{5 -- 78$\times$} faster; and 2D convolution runs only 1.4 -- 1.9$\times$ faster. The relatively low performance of 2D convolution and, to a lesser extent, Matrix Max Pool is mainly due to their highly repetitive use of scalar arithmetic operations to manage data pointers. This greatly limits performance despite both benchmarks using optimized vector reduction and dot product functions, respectively. We believe that strided vector memory operations can improve the performance of both applications, and we are currently working on further optimizing these benchmarks. It is also worth noting that vectorized benchmarks with larger data profiles achieve better performance than those with smaller data profiles. This is mainly due to vector overhead instructions (e.g. setting the vector length) having a lower impact on execution time when vector instructions operate on longer vectors.

\begin{table*}[ht!]
 \centering

\scalebox{0.8}{
\begin{tabular}{@{\extracolsep{2pt}}||l||c|c|c||c|c|c||c|c|c||}
\hline \hline
 {}  & \multicolumn{3}{c||}{Small Data Profile}  & \multicolumn{3}{c||}{Medium Data Profile} & \multicolumn{3}{c||}{Large Data Profile} \\
 \cline{2-4} 
 \cline{5-7} 
 \cline{8-10} 
 Operation  & Scalar (J) & Vector (J)  & Ratio & Scalar (J) & Vector (J)  & Ratio & Scalar (J)& Vector (J) & Ratio  \\ 
\hline
Vector Addition    & $8.6\times{}10^{-6} $  &  $1.4\times{}10^{-7} $  & 1.6\% & $6.8\times{}10^{-5} $   & $9.7\times{}10^{-7} $   & 1.4\% & $5.44\times{}10^{-4} $   & $7.6\times{}10^{-6} $ & 1.4\%  \\ \hline
Vector Multiplication    & $8.5\times{}10^{-6} $ & $1.3\times{}10^{-7} $  & 1.6\% & $6.9\times{}10^{-5} $  & $9.6\times{}10^{-7} $  & 1.4\% & $5.3\times{}10^{-4} $  & $7.5\times{}10^{-6} $ & 1.4\%   \\ \hline
Vector Dot Product    & $3.8\times{}10^{-6} $ & $1.7\times{}10^{-7} $  & 4.4\% & $3.0\times{}10^{-5} $  & $1.0\times{}10^{-6} $  & 3.4\% &  $2.4\times{}10^{-4} $  & $8.0\times{}10^{-6} $ & 3.3\% \\ \hline
  Vector Max Reduction  & $3.4\times{}10^{-6} $  & $1.1\times{}10^{-7} $  & 3.4\% &  $2.6\times{}10^{-5} $ & $6.1\times{}10^{-7} $ & 2.3\% & $2.1\times{}10^{-4} $  & $4.5\times{}10^{-6} $  & 2.1\%  \\ \hline
  Vector ReLu  & $3.5\times{}10^{-6} $  & $1.1\times{}10^{-7} $  & 3.2\% &  $2.8\times{}10^{-5} $ & $7.9\times{}10^{-7} $ & 2.9\% & $2.2\times{}10^{-4} $  & $6.2\times{}10^{-6} $ & 2.8\%  \\ \hline
Matrix Addition  & $5.52\times{}10^{-4} $ & $1.4\times{}10^{-5} $ &  2.5\% & $3.5\times{}10^{-2} $  & $5.4\times{}10^{-4} $ & 1.5\%  & $2.2\times{}10^{0} $  & $3.2\times{}10^{-2} $ &  1.4\% \\ \hline
Matrix Multiplication  & $3.0\times{}10^{-2} $  & $1.4\times{}10^{-3} $ & 4.6\% & $1.5\times{}10^{1} $  & $3.3\times{}10^{-1} $  & 2.2\% & $7.6\times{}10^{3} $  & $1.4\times{}10^{2} $ & 1.9\% \\ \hline
 Matrix Max Pool  & $9.2\times{}10^{-4} $  & $1.88\times{}10^{-4} $ & 20.5\% & $5.9\times{}10^{-2} $  & $1.2\times{}10^{-2} $  & 20.4\% & $3.8\times{}10^{0} $  & $7.65\times{}10^{-1} $ &  20.4\% \\ \hline
 2D Convolution  & $3.4\times{}10^{0} $  & $1.9\times{}10^{0} $ & 57.3\% & $4.5\times{}10^{0} $ & $3.2\times{}10^{0} $ & 70.4\% & $6.0\times{}10^{0} $ &  $6.7\times{}10^{0} $ & 79.9\% \\ \hline \hline
\end{tabular}}
\caption{Energy Consumption Analysis}
\label{tab:Energy_Analysis}
\end{table*}
   
Table~\ref{tab:Energy_Analysis} shows the energy consumed by the scalar and vectorized benchmarks for different data size profiles. The Table also shows the ratio of energy consumed by vectorized benchmarks relative to corresponding scalar benchmarks. Our results show that the Arrow accelerator achieves significant energy reductions relative to a~scalar processor. This is mainly due to the significant reduction in execution time, which directly impacts energy consumption. Comparing the energy consumed by the vectorized benchmarks to that consumed by the scalar benchmarks we find that the vector benchmarks consume 96\% to 99\% less energy; the matrix benchmarks consume 80\% to 99\% less energy; and 2D convolution consumes 20\% to 43\% less energy. While still significant, we believe that further optimizations to 2D convolution and Matrix Max Pool will result in further reductions in energy consumption.

These results highlight the performance and energy advantages of the RVV-based Arrow architecture, and validate its suitability for edge machine learning inference applications.

\section{Conclusions and Future Work} \label{sec:conc}

In this paper we described the architecture of our RVV-based Arrow accelerator and evaluated its performance and energy consumption. Our results confirm its suitability for edge machine learning inference applications. We plan to extend this work by evaluating the Arrow accelerator when it is more tightly integrated into the datapath of a RISC-V processor like the SweRV EH1. We also plan to develop a gem5 simulation model to perform a more thorough performance study of the Arrow accelerator using the MLPerf and TinyMLPerf benchmark suites. Finally, we plan to expand the Arrow architecture to support ML-friendly arithmetic operations and data types like bfloat16 and posits.

\bibliographystyle{ACM-Reference-Format}
\bibliography{references}

\end{document}